%% file: rootV3_finalSubmit.tex
\documentclass[letterpaper, 10 pt, conference]{ieeeconf}  

\IEEEoverridecommandlockouts                              
\usepackage{graphics} 
\usepackage{epsfig} 
\usepackage{mathptmx} 
\DeclareMathAlphabet{\mathcal}{OMS}{cmsy}{m}{n}
\usepackage{times} 
\usepackage{amsmath} 
\usepackage{amssymb}  

\let\labelindent\relax
\usepackage{enumitem}
\usepackage{lipsum}
\usepackage{tikz}
\usetikzlibrary{plotmarks}
\usepackage{pgfplots}
\pgfplotsset{compat=1.18}

\input{Custom_commands.tex}

\usepackage{xcolor}
\input{Colors.tex}
\usepackage{algorithm}
\usepackage{algpseudocode}
\usepackage{ntheorem}
\newtheorem*{remark}{Remark}

\title{\LARGE \bf
Automatic Basis Function Selection in Iterative Learning Control: \\
A Sparsity-Promoting Approach Applied to an Industrial Printer
}

\author{Tjeerd Ickenroth$^{1}$, Max van Haren$^{1}$, Johan Kon$^{1}$, Max van Meer$^{1}$, Jilles van Hulst$^{1}$, and Tom Oomen$^{1,2}$
\thanks{$^1$Tjeerd Ickenroth (corresponding author), Max van Haren, Johan Kon, Max van Meer, Jilles van Hulst, and Tom Oomen are with the Control Systems Technology Section, department of Mechanical Engineering, Eindhoven University of Technology, Eindhoven, The Netherlands.\newline
\tt\small t.s.ickenroth@tue.nl}
\thanks{$^2$Tom Oomen is with the Delft Center for Systems and Control, Delft University of Technology, Delft, The Netherlands.}
}

\begin{document}

\maketitle
\thispagestyle{empty}
\pagestyle{empty}

\begin{abstract}
Iterative learning control (ILC) techniques are capable of improving the tracking performance of control systems that repeatedly perform similar tasks by utilizing data from past iterations. 
The aim of this paper is to design a systematic approach for learning parameterized feedforward signals with limited complexity. The developed method involves an iterative learning control in conjunction with a data-driven sparse subset selection procedure for basis function selection.
The ILC algorithm that employs sparse optimization is able to automatically select relevant basis functions and is validated on an industrial flatbed printer.
\end{abstract}

\section{INTRODUCTION}
Learning in control systems allows to improve tracking performance for systems that need to perform multiple motion tasks. By effectively exploiting large amounts of data that is currently available control systems can achieve outstanding performance. The data contains valuable information about the to be controlled system that subsequently can be used to learn important system properties. These properties are key to be able to effectively control mechatronic systems up to high precision when performing multiple tasks.\\
%
Iterative Learning Control (ILC) has demonstrated considerable success in improving performance in systems that operate repetitively. By utilizing data from previous iterations, ILC refines the control input to effectively reject disturbances that repeat with each trial. ILC has been successfully implemented in real world applications in varying industries, e.g. robotics~\cite{Axehill2014}, mechatronics~\cite{Bolder2018Data-drivenFilters}~\cite{Huang2014High-performanceScheduling}, and manufacturing~\cite{Hoelzle2016}. In spite of excellent reported tracking accuracy, the learned control signal is optimal for a fixed reference trajectory only. Since extrapolation to other references can lead to significant accuracy degradations~\cite{Boeren2016Frequency-DomainEquipment}, it is essential that extrapolation capabilities are introduced in ILC to achieve improvements for multiple tasks.\\
%
Since ILC is traditionally non-robust to varying references, several directions have been explored to develop such ILC algorithms that are flexible to task variations.
Recent successful developments in ILC with basis functions include a physics based approach~\cite{Lambrechts2005TrajectorySystems}, Bayesian priors~\cite{Blanken2020}, polynomial basis functions~\cite{Bolder2014UsingPrinter}~\cite{VanDerMeulen2008FixedPrinter}~\cite{VanDeWijdeven2010UsingTheory}, and rational basis functions~\cite{Bolder2017}~\cite{VanZundert2016OptimalityTasks}, which all aim to approximate the inverse system. An alternative approach is shown in~\cite{Hoelzle2011BasisDeposition} that requires a predefined set of reference tasks.\\
%
The specification of basis functions on the basis of physical insight highly depends on the application at hand
and the quality of the prior, hence the performance of such approaches varies. In particular, the results in~\cite{Lambrechts2005TrajectorySystems}
assume a very specific system structure, which has shown exceptional results for varying references in industrial applications in case the system is in the model class. If the system is not in the model class, other
basis functions may be preferred, which heavily depend on the quality of the model of the system.\\
%
Although important developments have been made in task-flexible ILC and several successful applications have been reported, the basis function parameterization for performance and task flexibility is an open question. The aim of this paper is to perform automatic basis function selection for ILC.\\
%
Basis function selection for ILC involves several unique requirements, including task
flexibility, performance, and low-order parameterizations for implementation and noise
attenuation.
Sparse subset selection methods in dynamic systems have been investigated in the field of system identification~\cite{Rojas2011SparseCriterion}~\cite{Ljung2011FourIdentificationg}~\cite{Brunton2016SparseSINDYc} where it is a common problem to estimate an accurate mathematical model of a real system using as few parameters as possible. 
Many techniques for sparse estimation have been successfully used for model structure selection in linear regression settings. 
For example, in \textit{Forward Selection} regressors are added one by one according to how statistically significant they are~\cite{Ingene1981AppliedRegression}. 
Another approach to efficiently enforce sparsity is by the use of $\ell_1$-regularization, i.e., a penalty on the size of the parameter vector is added to the cost function and is referred as \textit{Least Absolute Shrinkage and Selection Operator} (LASSO)~\cite{Tibshirani1996RegressionLasso}.
\textit{Least Angle Regression and Shrinkage} (LARS) is able to efficiently solve a LASSO problem along the solution path~\cite{Efron2004LeastRegression}. First adaptation of sparsity in ILC has been investigated in~\cite{Oomen_Rojas_2017} where it has been used to construct sparse command signals. However, the use of such techniques to select basis functions is immature and not developed.\\
The main contribution of this paper is to develop a framework for automatic basis function selection based on sparse optimization for ILC, including experimental implementation and comparison.\\
The paper is structured as follows. The learning control problem is formally stated in Section \ref{sec:Problem_Formulation}. In Section \ref{sec:Method} the solution is proposed. Section \ref{sec:basis_function_design} elaborates on the parameterization of feedforward filters. In Section \ref{sec:Experimental_validation}, an experimental case study is presented that reveals the advantages of automatic basis function in ILC.
%
%
\\
\textit{Notation:} 
Throughout, $\|x\|_{\ell_p}$ denotes the usual $\ell_p$-norm, $p \in \mathbb{N}$. Also, $\|x\|_0 = \sum_i (x_i\neq0)$, i.e., the cardinality of $x$. Note that $\|x\|_0$ is not a norm, since it does not satisfy the homogeneity property. It relates to the general $\ell_p$-norm by considering the limit $p \rightarrow 0$ of $\|x\|_p$.
A vector $x \in \mathbb{R}^N$ is called sparse if many of its components are zero, in which case $\|x\|_0 \ll N$.
Let $\mathcal{H}(z)$ denote a discrete-time, linear time-invariant (LTI), single-input, single-output system. Vectors are denoted as lowercase letters and matrices as uppercase letters, e.g., $x$ and $X$. The columns and rows with index $i$ of a matrix $X$ are denoted by $\text{col}_i(X)$ and $\text{row}_i(X)$, respectively.
Let $h(k)~\forall k \in \mathbb{Z}$ be the impulse response coefficients of $\mathcal{H}(z)$, with the infinite impulse response $y(k) = \sum_{\tau=-\infty}^{\infty} h(\tau)u(k-\tau)$. Let $u(k) = 0$ for $k<0$ and $k\geq N$ to obtain the finite-time convolution
$$
\resizebox{\hsize}{!}{$
\underbrace{\left[\begin{array}{c}
y(0) \\
y(1) \\
\vdots \\
y(N-1)
\end{array}\right]}_y=\underbrace{\left[\begin{array}{cccc}
h(0) & h(-1) & \cdots & h(-N+1) \\
h(1) & h(0) & \cdots & h(-N+2) \\
\vdots & \vdots & \ddots & h(-1) \\
h(N-1) & h(N-2) & \cdots & h(0)
\end{array}\right]}_H \underbrace{\left[\begin{array}{c}
u(0) \\
u(1) \\
\vdots \\
u(N-1)
\end{array}\right]}_u,
$}
$$
with $y,~u \in \mathbb{R}^N$ and $H \in \mathbb{R}^{N\times N}$ is the finite-time convolution matrix corresponding to $\mathcal{H}(z)$.



\section{PROBLEM FORMULATION}\label{sec:Problem_Formulation}
In this section, the problem is formulated. First, the problem setup is presented. Second, the experimental setup is introduced. Finally, the problem that is addressed in this paper is defined.
\begin{figure}[b]
    \centering
    \includegraphics[width=0.8\linewidth]{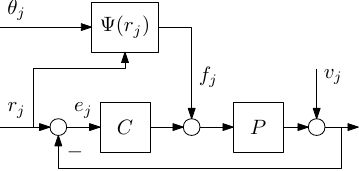}
    \caption{Control structure considered.}
    \label{fig:Control_scheme}
\end{figure}
\subsection{Control setup}
Consider the closed-loop system shown in Fig. \ref{fig:Control_scheme}. The system and stabilizing feedback controller are denoted by $P$ and $C$, respectively. Both are considered to be discrete time, single-input single-output (SISO), and linear time-invariant (LTI) systems.
The associated ILC system, in this so-called lifted notation, is
\begin{equation}\label{ILCsystem}
    e_j = S(r_j - Pf_j) - Sv_j,
\end{equation}
where $S = (I_N+PC)^{-1}$ is the sensitivity, $j \in \mathbb{N}$ the trial, or experiment, of length $N \in \mathbb{N}$, $e_j \in \mathbb{R}^N$ the error signal to be minimized, $r_j \in \mathbb{R}^N$ the reference signal that is possibly trial-varying, $f_j \in \mathbb{R}^N$ the command signal, and $v_j \in \mathbb{R}^N$ represents trial-varying disturbances, including measurement noise. \\
Similarly to \eqref{ILCsystem}, the error at trial $j+1$ is given by
\begin{equation}
    e_{j+1} = S(r_{j+1} - Pf_{j+1}) - Sv_{j+1},
\end{equation}
and under the assumption that the reference is trial invariant, i.e., $r_{j+1} = r_j$, the error propagation in lifted notation is given by
\begin{equation}\label{error_propagation}
    e_{j+1} = e_j - J(f_{j+1}-f_j) - S(v_{j+1}-v_j),
\end{equation}
where $J=PS \in \mathbb{R}^{N\times N}$ the finite-time convolution matrix of the process sensitivity.

\subsection{Application example}
\begin{figure}[t]
    \centering
    \includegraphics[width=\linewidth]{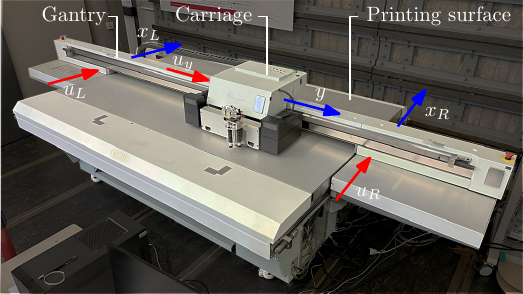}
    \caption{Experimental setup: Canon Production Printing Arizona 550 GT flatbed printing system, with the inputs indicated in red and outputs in blue.}
    \label{fig:Arizona}
\end{figure}

\begin{figure}[t]
    \centering
    \includegraphics[width=0.8\linewidth]{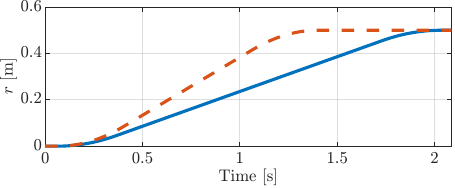}
    \caption{The first $\markerline{MatlabBlue}$ and second $\markerline{MatlabRed}[dashed]$ 
    faster motion tasks being applied to the Arizona flatbed printer.}
    \label{fig:reference}
\end{figure}
The mechatronic system considered in this paper is the flatbed printer shown in Fig. \ref{fig:Arizona}. In contrast to standard consumer printers, the medium is fixed on the printing surface. The carriage, which contains the print heads, translates in $y$-direction along the gantry, which translates in $x$ and rotates in $R_z$. Hence, the printer is able to hover over the printing surface in both $x$ and $y$ directions. In this work a single-input single-output situation is considered, in which only the carriage is moved.\\
The printer must be able to perform multiple print tasks on sheets of varying widths and varying materials with high precision. The duration of a print task depends on the material to be printed and the type of print, e.g., glossy or matt. Therefore, the motion tasks that are applied to the carriage are not one single non-changing reference, but a set of references that require high accuracy tracking. Two examples of references with different printing speed and carriage acceleration are shown in Fig. \ref{fig:reference}.

\subsection{Learning problem}\label{sec:Learning problem}
To allow flexibility in the motion task, the feedforward signal $f_j$ is parameterized by a set of basis functions, denoted by $\Psi(r_j) \in \mathbb{R}^{N\times N_\theta}$, and learning parameters $\theta_j \in \mathbb{R}^{N_\theta}$, resulting in
\begin{equation}\label{ffw_parameterization}
    f_j = \Psi(r_j)\theta_j,
\end{equation}
as shown in Fig. \ref{fig:Control_scheme}.
Substituting \eqref{ffw_parameterization} into \eqref{error_propagation} the error propagation becomes
\begin{equation}\label{error_propagation_new}
    e_{j+1} = e_j - J\Psi(r)(\theta_{j+1}-\theta_j) - S(v_{j+1}-v_j),
\end{equation}
which shows that $e_{j+1}$ is linear in the parameters $\theta_{j+1}$. The objective of feedforward control in ILC is to minimize the tracking error of the next trial, $e_{j+1}$ in \eqref{error_propagation_new}, by designing $\theta_{j+1}$ based on measured data of the previous trial, i.e., $e_j$ and $\theta_j$.\\
The parameters $\theta_j$, along with the basis $\Psi(r_j)$, aim to describe the inverse system $P^{-1}$, i.e., $\Psi(r_j)\theta_j \approx P^{-1}r_j$ to obtain zero tracking error in the absence of noise, see \eqref{ILCsystem}. However, the tracking performance is limited by the ability of $\Psi(r)\theta$ to describe the inverse system and heavily relies on the selection of basis $\Psi(r)$. \\
Therefore, the problem addressed in this paper is to develop an ILC algorithm that learns the optimal learning parameters $\theta_j$ while at the same time automatically selects a subset of basis functions $\Psi_\mathcal{S}(r) \subseteq \Psi(r)$ that contribute to high tracking performance. This is summarized in the following requirements:
%
\begin{enumerate}[label=R\arabic*)]
    \item the iterations \eqref{error_propagation_new} lead to a small error $e_j$ in the presence of both trial-varying reference $r_j$ and disturbances $v_j$;
    \item a rich basis $\Psi(r) \in \mathbb{R}^{N\times N_\theta}$ can be posed, i.e., $N_\theta$ can be large, whereas the algorithm automatically selects a (sparse) subset $\Psi_\mathcal{S}(r) \in \mathbb{R}^{N\times n_\theta} \subseteq \Psi(r)$ with $n_\theta \ll N_\theta$ parameters, thereby controlling the computational load of the algorithm and be robust to trial-varying effects.
\end{enumerate}

\section{SPARSE BASIS FUNCTION SELECTION}\label{sec:Method}
%
In this section the Sparse Basis Function Iterative Learning Control (SBF-ILC) algorithm is provided that satisfies Requirements R1 and R2. First, it is shown how to enforce exact sparsity in the learning parameters $\theta$. Second, the SBF-ILC algorithm is provided.

\subsection{Enforcing exact sparsity in the learning parameters}
\begin{figure}[b]
    \centering
    \includegraphics[width=0.5\linewidth]{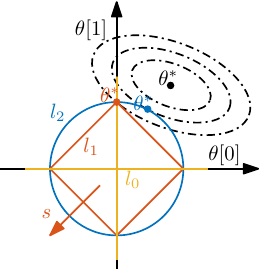}
    \caption{Enforcing sparsity in ILC. Assuming $N_\theta=2$, hence $\theta$ contains two elements. The ellipsoidal contour lines of objective \eqref{Unconstrained_cost} are plotted in $\markerline{MatlabBlack}[dash dot]$. The $\ell_0$ constraint set is plotted in $\markerline{MatlabYellow}$, which is, obviously, non-convex. The $\ell_1$ constraint set is plotted in $\markerline{MatlabRed}$. The $\ell_2$ constraint set is plotted in $\markerline{MatlabBlue}$. $\theta^*$ denotes the unconstrained optimal solution to objective function \eqref{Unconstrained_cost}. The optimal constrained solution is found at the point where the contour line of the objective function first touches the constraint set, for the $\ell_1$-case, this implies that $\theta[0] = 0$, hence $\color{MatlabRed}{\theta^*}$ is sparse. In contrast, for the ridge regression including the $\ell_2$ constraint, the solution $\color{MatlabBlue}{\theta^*}$ is not sparse.}
    \label{fig:cost_and_lars_path}
\end{figure}

The objective of feedforward control in ILC is to minimize the tracking error of the next trial, $e_{j+1}$ in \eqref{error_propagation_new}, by learning the parameters $\theta_{j+1}$ based on measured data of the previous trial for a given set of basis functions $\Psi(r_j)$.
The optimization problem is defined as
\begin{equation}\label{Unconstrained_cost}
\begin{split}
    \mathcal{J}(\theta_{j+1}) = \frac{1}{2}\| e_{j+1}(\theta_{j+1}) \|_{W_e}^2 + \frac{1}{2}\| \Psi(r_j)\theta_{j+1} \|_{W_f}^2\\
    + \frac{1}{2}\| \Psi(r_j)(\theta_{j+1} - \theta_j) \|_{W_{\Delta f}}^2,
\end{split}
\end{equation}
where $\|x\|_W^2 = x^\top Wx$, and $W_e,~W_f,~W_{\Delta f} \in \mathbb{R}^{N\times N}$ are symmetric positive (semi-)definite weighting matrices.
Problem \eqref{Unconstrained_cost} is quadratic in the optimization variables $\theta_{j+1}$, and hence, has a minimizer that is analytically computed as
\begin{gather}\label{analytic_solution}
    \theta_{j+1} = Q\theta_j + Le_j, \quad \text{with}\\
    \begin{aligned}
    L &= \left(\Psi^\top (J^\top W_e J + W_f + W_{\Delta f})\Psi \right)^{-1} \Psi^\top J^\top W_e,\\
    Q &= \left(\Psi^\top (J^\top W_e J + W_f + W_{\Delta f})\Psi \right)^{-1} \Psi^\top (J^\top W_e J + W_{\Delta f})\Psi, \nonumber
    \end{aligned}
\end{gather}
with learning and robustness matrices $L \in \mathbb{R}^{N_\theta \times N}$ and $Q \in \mathbb{R}^{N_\theta \times N_\theta}$, respectively. Here the reference dependency has been removed for brevity.\\
In accordance with requirements R1 and R2, and given a large set of basis functions $\Psi(r_j) \in \mathbb{R}^{N\times N_\theta}$, a (sparse) subset 
$$\Psi_\mathcal{S}(r_j) \in \mathbb{R}^{N\times n_\theta} \subseteq \Psi(r_j)$$
with $n_\theta$ learning parameters need to be automatically selected to control the computational load that at the same time leads to small tracking error $e_j$. 
Sparsity of a vector is measured directly through the amount of nonzero elements in this vector, i.e., $\|\theta_{j+1}\|_0$. The addition of the $\|\theta_{j+1}\|_0$ constraint in \eqref{Unconstrained_cost} results, however, in a non-convex problem, which in fact is NP-hard~\cite{Natarajan1995SparseSystems}. 
To overcome this, the problem is relaxed by an $\ell_1$-norm that promotes sparsity in the parameters $\theta_{j+1}$, generally referred as LASSO, and is given by the following optimization problem
\begin{equation}\label{lasso_contrained}
        \min_{\theta_{j+1}} \quad \mathcal{J}(\theta_{j+1}) \quad \text{s.t.} \quad \|\theta_{j+1}\|_1 \leq s,
\end{equation}
where $s \in \Re_{>0}$ is an upper bound on the $\ell_1$-norm of the parameters: a small $s$ leads to highly sparse results. The Lagrangian of \eqref{lasso_contrained} is the non-smooth objective
\begin{equation}\label{lasso_regularized}
    \min_{\theta_{j+1}} \quad \mathcal{J}(\theta_{j+1}) + \lambda\|\theta_{j+1}\|_1
\end{equation}
where $\lambda > 0$ can be interpreted as a regularization parameter: a large penalty $\lambda$ corresponds to a small bound $s$, and vice versa.\\
The level of sparsity is controlled by the value of $s$ in \eqref{lasso_contrained} (or $\lambda$ in \eqref{lasso_regularized}) which is illustrated in Fig. \ref{fig:cost_and_lars_path}. 
Here, the contours of the unconstrained objective \eqref{Unconstrained_cost}, along with the contours of the $\ell_0$, $\ell_1$, and $\ell_2$ constraint sets are shown. 
It is geometrically clear that the $\ell_0$ constraint defines a non-convex set which is challenging to solve, and that the $\ell_1$ constraint promotes a sparse solution.
\\
Exact sparsity in the learning parameters can be obtained by solving regularization problem \eqref{lasso_regularized} along the regularization path via Least Angle Regression and Shrinkage \cite{Efron2004LeastRegression}.
Shrinkage refers to the process where the selected parameters are systematically reduced in magnitude, resulting in biased estimates of the obtained learning parameters. Hence, it is crucial to take one additional debiasing step to obtain the true learning parameters as shown in \cite{Oomen_Rojas_2017}.

\subsection{SBF-ILC algorithm}
\begin{algorithm}[t]
    \caption{Sparse Basis Function ILC}\label{alg:LARS}
    \begin{algorithmic}[1]
\State Initialize diagonal weighting matrices $W_e,~W_f,~W_{\Delta f} \succcurlyeq 0$,
\State Initialize number of parameters $N_\theta$ and cardinality $n_\theta$,
\State Construct $\Psi(r_j)$ and $X$ (see \eqref{X_and_Y}),
\State Initialize parameters $\theta_{j \in 1,\dots,N_{\mathrm{trials}}} = \textbf{0}^{\textcolor{black}{N_\theta \times 1}}$.
\For {$j \in 1,\dots,N_{\mathrm{trials}}$}
\State Update feedforward $f_j = \Psi(r_j)\theta_j$;
\State Run an experiment $e_j = Sr_j - Jf_j$;
\State Construct $Y_j$ (see \eqref{X_and_Y});
\State Find \textcolor{black}{sparse vector} $\tilde{\theta}_{j+1}$ = Lars($X,Y_j,n_\theta$) (\textit{biased}).
\Statex Debiasing: 
\State Find \textcolor{black}{$n_\theta$} non-sparse indices $\tilde{\theta}_{j+1}[i] \neq 0$;
\State Construct active set $X_A = \text{col}_i (X) \subseteq X$;
\State Calculate the least squares solution $\text{row}_i (\theta_{j+1}) = (X_A^\top X_A)^{-1}X_A^\top Y_j$ (\textit{debiased}).
\EndFor
    \end{algorithmic}
\end{algorithm}
The SBF-ILC algorithm is capable of efficiently selecting a (sparse) subset of basis functions, i.e. $\Psi_\mathcal{S}(r) \subseteq \Psi(r)$ with corresponding learning parameters $\theta$ that result in the lowest tracking error. 
The SBF-ILC algorithm that satisfies requirements R1 and R2 is provided in Algorithm \ref{alg:LARS}. Background on a LARS implementation can be retrieved from~\cite{Sjostrand2018SpaSM:Modeling}.\\ 
The expressions for the response vector $Y_j \in \mathbb{R^N}$ and predictor matrix $X \in \mathbb{R^{N\times N_\theta}}$ are obtained by rewriting the objective in \eqref{lasso_regularized} as
\begin{equation*}
    \| Y_j - X\theta_{j+1} \|_2^2 + \lambda\| \theta_{j+1} \|_1.
\end{equation*}
Substituting $e_{j+1}(\theta_{j+1}) = e_j - J\Psi(r_j)(\theta_{j+1}-\theta_j)$ in \eqref{lasso_regularized} and using that $\|.\|_W^2 = \|\sqrt{W}\|_2^2$ for $W$ diagonal, the matrices  $Y_j$ and $X$ are as follows 
\begin{equation}\label{X_and_Y}
        Y_j = \begin{bmatrix}
            \sqrt{W_e}(e_j + J\Psi(r_j)\theta_j)\\
            0_{N\times 1}\\
            -\sqrt{W_{\Delta f}} \Psi(r_j)\theta_j
        \end{bmatrix}, \quad
        X =
        \begin{bmatrix}
            \sqrt{W_e}J\\
            -\sqrt{W_f}\\
            -\sqrt{W_{\Delta f}}
        \end{bmatrix} \Psi(r_j).
\end{equation}
\begin{remark}
    Note that the SPILC algorithm in~\cite{Oomen_Rojas_2017} is recovered as a special case where $\Psi(r) = I_N$ to obtain sparse command signals instead of sparse basis function selection.
\end{remark}

\section{BASIS FUNCTION DESIGN}\label{sec:basis_function_design}
This section provides a possible parameterization for the set of basis functions. First, a parameterization based on physics is depicted that is often used in industry. Second, a parameterization of non-causal basis functions is introduced that allows to parameterize a large set of functions.\\
To be able to achieve good tracking performance, it is required to learn an accurate estimate of the inverse plant $P^{-1}$. The fit accuracy is determined by the selected set of basis functions $\Psi(r)$ together with their corresponding learning parameters $\theta$. The choice for basis $\Psi(r)$ has major influence on the fitting capabilities and is therefore extremely important.
As mentioned in Section \ref{sec:Learning problem} the basis functions with learning parameters aim to describe $\Psi(r_j)\theta_j \approx P^{-1}r_j$ to obtain zero tracking error. Hence, the set of basis functions as input to the SBF-ILC algorithm must be rich to explore the full potential of the approach.

\subsection{Physical basis functions}
In industry, basis functions are often based on physical insight on the system. The basis contains three physical interpretable parameters covering viscous friction, mass- and snap feedforward, i.e., 
\begin{equation}\label{IndustryBF}
    \Psi(r) = \begin{bmatrix}
    v & a & s
\end{bmatrix},
\end{equation}
where $v$ the velocity, $a$ the acceleration, and $s$ the snap~\cite{Lambrechts2005TrajectorySystems}. This parameterization leads to desirable applicable feedforward inputs but due to the limited amount of parameters in these basis functions the capability of approximating the inverse system remains limited, especially for complex mechatronic systems, resulting in unsatisfactory tracking performance.

\subsection{Non-causal basis functions}
Under the assumption that the system under consideration, i.e., $P$, is linear, it is known that a finite impulse response (FIR) filter is able to accurately describe the non-causal inverse system \cite{Blanken2020}. A non-causal FIR filter is described by
\begin{equation}\label{FIR_convolution}
    f_j[k] = \sum_{i=-n_p}^{N_\theta-n_p-1} \theta_j[i]r[k-i],
\end{equation}
where $f_j[k]$ the feedforward signal at trial $j$ at time instant $k$, and $n_p$ the number of non-causal samples.
Under the assumption that $r[k]=0$ for $k<0$, and $n_p=0$, then the expression in \eqref{FIR_convolution} is written in lifted notation as
\begin{equation}\label{FIR_basis_matrix}
\resizebox{.9\hsize}{!}{$
    f_j = 
    \begin{bmatrix}
        f_j[0]\\
        \vdots\\
        \vdots\\
        \vdots\\
        \vdots\\
        f_j[N-1]
    \end{bmatrix}
    =
    \underbrace{\begin{bmatrix}
        r[0] & 0 & \ldots & 0\\
        \vdots & \ddots & \ddots &\vdots\\
        \vdots & & \ddots& 0\\
        \vdots & & & r[0]\\
        \vdots & & & \vdots\\
        r[N-1] & \ldots & \ldots & r[N-N_\theta]
    \end{bmatrix}}_{\Psi_{\mathrm{FIR}}(r)}
    \underbrace{\begin{bmatrix}
        \theta_j[0]\\
        \vdots\\
        \vdots\\
        \vdots\\
        \vdots\\
        \theta_j[N_{\theta}-1]
    \end{bmatrix}}_{\theta_j}$},
\end{equation}
with $\Psi_{\mathrm{FIR}}(r)$ the toeplitz basis based on the finite impulse response. This FIR filter feedforward is able to outperform basis function ILC because of the better fitting capabilities of the inverse system due to an increased number of parameters.\\
The hyperparameters of a FIR are the total number of parameters $N_\theta$ and the number of non-causal parameters $n_p$ and require a precise selection.
For a large number of parameters $N_\theta$ and the presence of noise in the error signal $e_j$, this structure is sensitive to overfitting measurement noise which is a common problem in ILC~\cite{Oomen_Rojas_2017}.\\
The SBF-ILC algorithm has automatic selection until $n_\theta$ parameters have been found. Hence, it is allowed to select a wide range in causal and non-causal samples and it is only required to provide the total number of parameters $n_\theta$ that need to be selected.
%

%
\begin{remark}
    Note that an alternative parameterization for the set of basis functions can be formulated, such as an extension to basis functions which are nonlinear in the reference. The SBF-ILC algorithm would still perform an automatic subset selection.
\end{remark}

\section{EXPERIMENTAL VALIDATION}\label{sec:Experimental_validation}

In this section the proposed SBF-ILC algorithm is demonstrated using an industrial flatbed printer. First, the experimental setup is introduced. Second, the classes of ILC are presented together with the experimental outline. Last, the experimental results for all classes are presented.

\subsection{Experimental setup}
The proposed sparse ILC algorithm is validated using the industrial Arizona flatbed printer, shown in Fig. \ref{fig:Arizona}. The input $u_y$ is the force acting on the carriage, and the output $y$ is its measured position. The translation and rotation of the gantry have a reference equal to zero, and no feedforward is applied in these directions. 

\subsection{Experimental approach}
To validate robustness to trial-varying tasks, hence, satisfying requirement R1, two fourth-order polynomial reference signals are designed using the approach in~\cite{Lambrechts2005TrajectorySystems} with $N=2088$ samples and are shown in Fig. \ref{fig:reference}. The first reference is applied for 10 consecutive trials and is then switched to the second reference for another 10 consecutive trials. \\
To demonstrate the working principle of the SBF-ILC algorithm, hence satisfying requirement R2, the set of basis functions $\Psi(r_j)$ are constructed based on a FIR filter, see \eqref{FIR_basis_matrix}, where the total number of parameters $N_\theta = 200$ with $n_p=6$ samples preview. To validate performance of the algorithm it is compared to three different ILC methods that are generally applied. An overview of the experiments are listed below.

\begin{enumerate}[label=E\arabic*)]
\item \textit{SBF-ILC:} for $\Psi = \Psi_{\mathrm{FIR}}$, as in \eqref{FIR_basis_matrix}, where $N_\theta = 200$ and $n_p=6$. Experiments have been performed for increasing values of desired sparsity, i.e., number of nonzero parameters $n_\theta$, and is ranging from $n_\theta \in [1,~12]$.

\item \textit{Basis function (BF) ILC:} as is often considered in industry. Hence, the basis contains three physical parameters, i.e., $N_\theta=3$, and is equal to \eqref{IndustryBF}.

\item \textit{Norm optimal (NO) FIR:} with $\Psi = \Psi_{\mathrm{FIR}}$, as in \eqref{FIR_basis_matrix}, where $n_p=6$ and for two different number of parameters:
\begin{enumerate}
    \item $N_\theta=100$,
    \item $N_\theta=200$.
\end{enumerate}

\item \textit{Norm optimal (NO) ILC:} where $\Psi=I_N$ in \eqref{Unconstrained_cost}, hence, $N_\theta=N=2088$.
\end{enumerate}
In all experiments the diagonal weighting matrices $W_e,~W_f,$ and $W_{\Delta f}$ were held constant to ensure a fair comparison. Besides this, a compensation has been done to overcome the nonlinear effect of static friction, and is applied the same way for all experiments, hence it does not result in any advantage for either approach.

\subsection{Validation results}
\begin{table}[b]
\centering
\caption{Performance loss factors after a reference change.}
\label{tab:results}
\begin{tabular}{c|c|c|c|c}
 & \textbf{SBF-ILC} & \textbf{BF-ILC} & \textbf{NO FIR} & \textbf{NO ILC} \\ \hline
Parameters $n_\theta$ & 9 & 3 & 200 & 2088 \\ \hline
\begin{tabular}[c]{@{}c@{}}$\|e\|_2$ loss factor\\ at $r$-switch\end{tabular} & 1.67 & 2.32 & 1.72 & 42.66
\end{tabular}
\end{table}
\begin{figure}[b!]
    \centering
    \includegraphics[width=\linewidth]{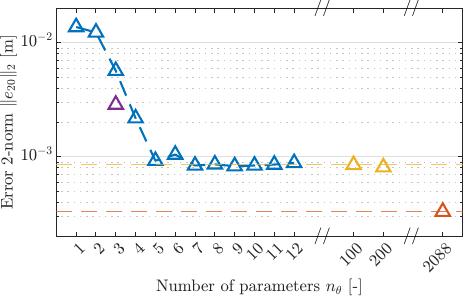}
    \caption{Experimental error 2-norm at trial $j=20$ for basis function ILC with $N_\theta=3$ in $\marker{MatlabPurple}[triangle]$, SBF-ILC for $n_\theta \in [1,~12]$ in $\markerline{MatlabBlue}[dashed][triangle]$, NO FIR for $N_\theta = \{100,~200\}$ in $\marker{MatlabYellow}[triangle]$, and NO ILC with $N_\theta=2088$ in $\marker{MatlabRed}[triangle]$. Indeed, increasing the number of parameters results in better performance up to an extent, as after $n_\theta=7$ the performance saturates.}
    \label{fig:Comparison_parameters}
\end{figure}
The error 2-norm for all experiments E1-E4 at the last trial $j=20$ are shown in Fig. \ref{fig:Comparison_parameters}. Table \ref{tab:results} shows the performance loss factor after switching reference at trial $j=11$. In Fig. \ref{fig:feedforward} the resulting command signal $f_j$ is shown at the final trial using SBF-ILC with $n_\theta=9$ and when using a FIR basis with $N_\theta=200$.\\
The following observations are made:
\begin{figure}[t]
    \centering
    \includegraphics[width=\linewidth]{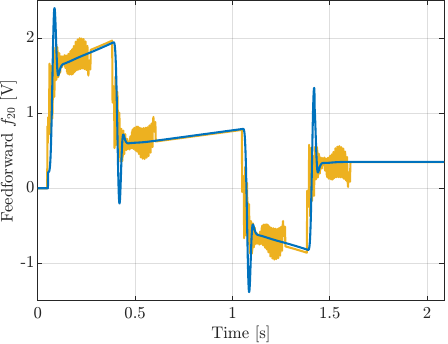}
    \caption{Experimentally learned feedforward at trial $j=20$ for NO FIR with $N_\theta=200$ parameters in $\markerline{MatlabYellow}$, and for SBF-ILC with $n_\theta=9$ parameters in $\markerline{MatlabBlue}$. SBF-ILC learns a smooth command signal using significantly less parameters compared to NO FIR, while both obtain the same performance.}
    \label{fig:feedforward}
\end{figure}
\begin{itemize}
    \item Fig. \ref{fig:Comparison_parameters} shows that SBF-ILC outperforms basis functions ILC when using $n_\theta \geq 4$ parameters. Note that 4 FIR filter parameters are required to reconstruct the same basis that is used in E2. Besides this, the performance improvement saturates when increasing the number of parameters. The result shows that using SBF-ILC with $n_\theta=7$ parameters achieves the same performance when using FIR ILC with $N_\theta=200$ parameters. 
    \item Table \ref{tab:results} shows that the proposed SBF-ILC approach is robust to a reference change, since the loss factor is small compared to the other approaches.
    \item Fig. \ref{fig:Comparison_parameters} shows that SBF-ILC with only $n_\theta \in [7,~12]$ parameters obtains equal performance compared to FIR ILC having $N_\theta \in \{100,~200\}$ parameters. However, the command signals $f_j$ that are obtained after learning are dissimilar and are shown in Fig. \ref{fig:feedforward}. Here it is shown that the FIR feedforward contains high frequency content which is due to overfitting the parameters, whereas the SBF-ILC feedforward is a smooth signal. This makes the SBF-ILC result much more favorable to apply on a real mechatronic setup.
\end{itemize}

\section{CONCLUSION}\label{sec:Conclusion}
In this paper, a sparse basis function iterative learning control (SBF-ILC) algorithm has been developed, which integrates automatic basis function selection to achieve both high tracking performance and robustness to varying tasks. The algorithm accommodates a large set of candidate basis functions, from which a sparse subset is selected by solving a LASSO optimization problem. This selection process identifies the basis functions that contribute most to minimizing the tracking error, with the level of sparsity directly controlled by the user.
The SBF-ILC algorithm is validated on an industrial flatbed printer where it is compared to traditional basis function ILC, FIR ILC, and norm optimal ILC. Results demonstrate that SBF-ILC outperforms traditional basis function ILC, and can achieve equivalent performance to FIR ILC with only a fraction of the learning parameters. Consequently, SBF-ILC offers high tracking accuracy with reduced computational cost, producing smooth control signals suitable for industrial mechatronic applications.

\section*{ACKNOWLEDGEMENT}
This project is co-financed by Holland High Tech, top sector High-Tech Systems and Materials, with a PPP innovation subsidy for public-private partnerships for research and development.
The authors gratefully acknowledge the contributions to this paper through a challenge-based learning project by 
Tim Aarts, 
Remco Bertels, 
Matthijs van Brunschot, 
Armando Cerullo, 
Yuri Copal, 
Hein van Dal, 
Roel Drenth, 
Hessel van Gemert, 
Bas Klis, 
Olaf van Lamsweerde, 
Gijs van Meerbeeck, 
Tom Minten, 
Gert Vankan, and 
Teun Wijfjes.

\bibliographystyle{ieeetr}
\bibliography{references}

\end{document}

%% file: Custom_commands.tex
\DeclareDocumentCommand{\markerline}{m O{solid} O{} O{2.5pt} O{0.8pt} O{10pt} O{1} O{1}}{\raisebox{0.5ex} 
	\protect
	(\begin{tikzpicture}
		\protect
		\draw[color=#1,#2,line width=#5, draw opacity = #8] (0,0)--(#6,0);
		\protect
		\node[mark size=#4, color = #1, line width=#5, fill opacity = #7] at (0.5*#6,0) {\pgfuseplotmark{#3}};
\end{tikzpicture})}

\DeclareDocumentCommand{\marker}{m O{} O{2.5pt} O{0.8pt} O{1} O{1}}{\raisebox{0.5ex} 
	\protect
	(\begin{tikzpicture}
		\protect
		\node[mark size=#3, color = #1, line width=#4, fill opacity = #6] at (0,0) {\pgfuseplotmark{#2}};
\end{tikzpicture})}

\makeatletter
\renewcommand*\env@matrix[1][*\c@MaxMatrixCols c]{%
  \hskip -\arraycolsep
  \let\@ifnextchar\new@ifnextchar
  \array{#1}}
\makeatother

%% file: Colors.tex
\definecolor{MatlabBlue}{rgb}    {0     , 0.4470, 0.7410}
\definecolor{MatlabRed}{rgb}     {0.8500, 0.3250, 0.0980}
\definecolor{MatlabYellow}{rgb}  {0.9290, 0.6940, 0.1250}
\definecolor{MatlabPurple}{rgb}  {0.4940, 0.1840, 0.5560}
\definecolor{MatlabGreen}{rgb}   {0.4660, 0.6740, 0.1880}
\definecolor{MatlabBabyBlue}{rgb}{0.3010, 0.7450, 0.9330}
\definecolor{MatlabGray}{rgb}{0.5, 0.5, 0.5}
\definecolor{MatlabLightGray}{rgb}{0.75, 0.75, 0.75}
\definecolor{MatlabBlack}{rgb}{0, 0, 0}

\definecolor{MatlabLightGray4}{rgb}{0.875, 0.875, 0.875}
\definecolor{MatlabLightGray3}{rgb}{0.85, 0.85, 0.85}
\definecolor{MatlabLightGray2}{rgb}{0.775, 0.775, 0.775}
\definecolor{MatlabLightGray1}{rgb}{0.7, 0.7, 0.7}
\definecolor{MatlabGray30}{rgb}{0.3, 0.3, 0.3}
\definecolor{MatlabGray40}{rgb}{0.4, 0.4, 0.4}
\definecolor{MatlabGray50}{rgb}{0.5, 0.5, 0.5}
\definecolor{MatlabGray60}{rgb}{0.6, 0.6, 0.6}
\definecolor{MatlabGray70}{rgb}{0.7, 0.7, 0.7}
\definecolor{MatlabGray80}{rgb}{0.8, 0.8, 0.8}
\definecolor{MatlabGray90}{rgb}{0.9, 0.9, 0.9}